\documentclass[aps,prl,letterpaper,twocolumn,amsmath,amssymb,nofootinbib]{revtex4}
\usepackage{graphicx}% Include figure files
\newcommand{\Ochi}{\Omega_{\chi} {\rm h}^2}
\newcommand{\lp}{\left(}
\newcommand{\rp}{\right)}
\DeclareMathAlphabet   {\mathsc}{OT1}{cmr}{m}{sc}

\def\[{\left [}
\def\]{\right ]}
\def\({\left (}
\def\){\right )}
\newcommand{\lang}{\left\langle}
\newcommand{\rang}{\right\rangle}

\newcommand{\beq}{\begin{equation}}
\newcommand{\eeq}{\end{equation}}
\newcommand{\bea}{\begin{eqnarray}}
\newcommand{\eea}{\end{eqnarray}}
\newcommand{\GeV}      {~\mathrm{GeV}}
\newcommand{\TeV}      {~\mathrm{TeV}}

\newcommand{\UV}       {\mathsc{uv}}

\newcommand{\SUSY}     {\mathsc{susy}}

\newcommand{\order}{\mathcal{O}}

\newcommand{\gappeq}{\mathrel{\rlap {\raise.5ex\hbox{$>$}}
{\lower.5ex\hbox{$\sim$}}}}
\newcommand{\lappeq}{\mathrel{\rlap{\raise.5ex\hbox{$<$}}
{\lower.5ex\hbox{$\sim$}}}}

\begin{document}

%\preprint{MCTP-03-16}

\title{Non-Universal Gaugino Masses, CDMS, and the LHC}

\author{Michael Holmes}
\author{Brent D. Nelson}
\affiliation{Department of Physics, Northeastern University, Boston,
MA 02130, USA}

\date{\today}

\begin{abstract}
We consider the possibility that the recently reported events at the
CDMS-II direct dark matter detection experiment are the result of
coherent scattering of supersymmetric neutralinos. In such a
scenario we argue that non-universal soft supersymmetry breaking
gaugino masses are favored with a resulting lightest neutralino with
significant Higgsino and wino components. We discuss the
accompanying signals which must be seen at liquid-xenon direct
detection experiments and indirect detection experiments if such a
supersymmetric interpretation is to be maintained. We illustrate the
possible consequences for early discovery channels at the LHC via a
set of benchmark points designed to give rise to an observed event
rate comparable to the reported CDMS-II data.
\end{abstract}

%\pacs{14.80.Cp,12.60.Jv}
%\keywords{}

\maketitle

%%%%%%%%%%%%%%%%%%%%%%%%%%%%%%%%%%%%%%%%%%%%%%%%%%%%%%%%%%%%%%%%%%
{\em Introduction.} It has been widely held for some time that one
of the most attractive virtues of supersymmetry as a theory relevant
at the electroweak scale is the possible presence of a stable,
massive, neutral state in the superpartner spectrum capable of
explaining the missing non-baryonic matter in the
cosmos~\cite{Goldberg:1983nd}. An excellent candidate for this dark
matter particle is the lightest neutralino of
supersymmetry~\cite{Jungman:1995df,Bertone:2004pz}.
The diagrams which give rise to annihilation among relic neutralinos
are among the same diagrams which determine the interaction rate of
relic neutralinos in our local halo with terrestrial detectors.
Therefore, supersymmetric models whose lightest supersymmetric
particle (LSP) has the appropriate properties to account for the
needed non-baryonic dark matter are often also models which could
give rise to an observable signal in direct detection experiments.
Establishing the existence of such a dark matter candidate via the
observation of elastic scattering of the state off of target nuclei
has been the pursuit of a variety of experiments for some period of
time~\cite{Gaitskell:2004gd}.

Recently, the CDMS-II collaboration has reported the observation
of~2 nuclear recoil events consistent with the scattering of a
weakly-interacting massive particle (WIMP) from germanium nuclei in
a total exposure of 612~kg-days~\cite{Ahmed:2009zw}. The measured
recoil energies for these two events were 12.3 keV and 15.5 keV,
%with a third event that was not included in the signal sample
near the low end of the recoil energies considered. In this energy
range the CDMS collaboration estimates their signal efficiency to be
no less than 25\%, reaching a maximum of 32\% at 20~keV recoil
energies. The expected number of background events from
mis-identified electron scattering events or neutron scattering
events generated by radioactive decays and/or cosmic ray events was
estimated to be $\Delta N = 0.6 \pm 0.1\,{\rm (stat)}$ prior to
unblinding. This estimate was subsequently estimated to be $\Delta N
= 0.8 \pm 0.1\,{\rm (stat)} \pm 0.2\,{\rm (syst)}$ after
investigating the signal events in greater detail. It is therefore
quite possible that these events are simply mis-identified
background events. Taken as a signal they imply a coherent
scattering cross-section for a WIMP in the local halo at roughly the
few $\times 10^{-44}$ cm$^2$ level.

If these events are interpreted as the scattering of a
supersymmetric neutralino from the germanium components in the CDMS
detector a number of very exciting corollaries may follow. The
implications of these events has already been studied in a number of
interesting model
contexts~\cite{Kadastik:2009gx,Bottino:2009km,Feldman:2009pn,Ibe:2009pq}.
In this work we will outline some of the possible implications of
the CDMS data on the nature of supersymmetric models generally. To
illustrate these features we will consider a number of benchmark
points -- all of which could generate signals of the sort reported
by the CDMS experiment. One of the most important conclusions is
that some amount of non-universality in the gaugino soft
supersymmetry-breaking masses may need to be
present. We will discuss the crucial cross-checks and corroborating
observations that must be seen in the next round of data-taking in
both germanium- and xenon-based detectors for these conclusions to
hold. We will also consider the prospects for indirect detection via
the observation of the decay products from neutralino
pair-annihilation at gamma-ray observatories and the neutrino
observatory at IceCube. We will comment on the compatibility of any
supersymmetric interpretation of the CDMS data with recent
measurements of cosmic anti-matter fluxes from the PAMELA experiment
and with thermal calculations of the relic density of stable
neutralinos. Finally, we will briefly comment on what we may expect
from the LHC at center-of-mass energies of~7, 10~and 14~TeV for our
candidate models.\\

%%%%%%%%%%%%%%%%%%%%%%%%%%%%%%%%%%%%%%%%%%%%%%%%%%%%%%%%%%%%%%%%%%
{\em Direct Detection of Neutralino Recoils in Germanium.}

We begin with the methodology employed to translate the parameters
of supersymmetric models into predictions of event rates at the
CDMS-II in a given amount of exposure (fiducial volume $\times$ time
of data collection).
To compute the interaction rate of relic neutralinos with the nuclei
of the target material one considers both spin-dependent (SD) and
spin-independent (SI) interactions. For target nuclei with large
atomic numbers the SI interaction, which is coherent across all of
the nucleons in the nucleus, tends to dominate. The SI cross section
$\sigma^{\mathrm{SI}}$ is computed in {\tt
DarkSUSY}~\cite{Gondolo:2004sc} on an arbitrary nuclear target via
\begin{equation}\label{ds30}
    \sigma^{\mathrm{SI}}_{\chi i}=\frac{\mu^2_{i\chi}}{\pi}\big|
    ZG^p_s+(A-Z)G^n_s \big|^2\, ,
\end{equation}
where $i$ labels the nuclear species in the detector with nuclear
mass $M_i$, $\mu_{i\chi}$ is the reduced mass of the
nucleus/neutralino system $\mu_{i\chi}=m_{\chi} M_i/(m_\chi + M_i)$,
and $A$ and $Z$ are the target nucleus mass number and atomic
number, respectively. The quantities $G^p_s$ and $G^n_s$ represent
scalar four-fermion couplings of the neutralino to point-like
protons and neutrons. They can be described schematically as
\begin{equation}
G^N_s=\sum_{q=u,d,s,c,b,t}\langle N|\bar q q | N\rangle \times \(
{\rm SUSY\,\,parameters} \)\, , \end{equation}
where the quantity in parenthesis is calculable once the details of
the supersymmetric model are specified. The initial nuclear matrix
elements, however, are at present not calculable from first
principles. Their values must be inferred from pion-nucleon
scattering data. Depending on the methodology employed in this
analysis, different values for this important set of parameters can
be extracted -- particularly for the case of the $\pi\,N$
$\Sigma$-term~\cite{Bottino:1999ei,Bottino:2001dj,Pavan:2001wz}. The
importance of the resulting uncertainty in this parameter on
predictions for dark matter interaction cross-sections was recently
considered in~\cite{Ellis:2005mb,Ellis:2008hf,Bottino:2008mf}, where
it was shown to be potentially quite large. We will return to this
issue at the very end of this section. For what follows we will
simply use the default values in {\tt DarkSUSY~5.0.4} for all
nuclear matrix elements.

The differential rate of neutralino-nucleon scattering events (in
units of events/kg/day) can be computed from~(\ref{ds30}) according
to
\begin{equation}\label{ds29}
\frac{dR}{dE}=\sum_i c_i \frac{\rho_\chi \sigma_{\chi
i}|F_i(q_i)|^2} {2 m_\chi \mu^2_{i
\chi}}\int_{v_{min}}^\infty\frac{f(\vec v,t)}{v}d^3v \, ,
\end{equation}
where we sum over all nuclear species present, with $c_i$ being the
mass fraction of species $i$ in the detector. The quantity $f(\vec
v,t)\,d^3v$ is the neutralino velocity distribution (presumed to be
Maxwellian) with $v=|\vec v|$ the neutralino velocity relative to
the detector~\cite{Jungman:1995df}.
Finally $|F_i(q_i)|^2$ is the nuclear form factor for species $i$,
with $q_i=\sqrt{2M_i E}$ being the momentum transfer for a nuclear
recoil with energy $E$. For the purpose of this analysis we will use
the output differential rates from {\tt DarkSUSY}, calculated
via~(\ref{ds29}), over a range of recoil energies relevant to the
desired experiment. For the CDMS experiment the reported energy
range considered was
\begin{equation} 10\,{\rm keV} \leq E_{\rm recoil} \leq
100\,{\rm keV}\, . \label{CDMSrange} \end{equation}
We therefore perform a numerical integration of~(\ref{ds29})
%\begin{equation}\label{ratecalc}
%    R=\int_{E_{\rm min}}^{E_{\rm max}} \frac{dR}{dE}\,dE \, ,
%\end{equation}
%%
by constructing an interpolating function for the differential rate
sampled in $1$~keV intervals.

The sensitivity of the CDMS-II experiment, when translated into the
size of the SI cross section $\sigma^{\mathrm{SI}}_{\chi p}$, is
dependent on the mass of the LSP neutralino. For LSP masses in the
range $100\GeV \lappeq m_{\chi_1^0} \lappeq 300\GeV$ a detectable
event rate at CDMS-II implies a cross-section
$\sigma^{\mathrm{SI}}_{\chi p} \gappeq {\rm few} \times
10^{-44}\,{\rm cm}^2$. If we assume the signal efficiency is 30\%
for the CDMS-II experiment we have an effective exposure of
184~kg-days. For comparison to the data we note that a cross-section
$\sigma^{\mathrm{SI}}_{\chi p} = 10^{-44} \,{\rm cm}^2$ for
$m_{\chi_1^0} \simeq 150 \GeV$ implies approximately 0.5~expected
events in this effective exposure.

A scalar-scalar cross-section this large is not generic in
supersymmetric models, particularly those with universal gaugino
masses that tend to give rise to an overwhelmingly bino-like
LSP~\cite{Ellis:2005mb,Ellis:2001hv}. Values of
$\sigma^{\mathrm{SI}}_{\chi p}$ in the range for which CDMS-II is
sensitive tend to involve LSPs with a sizable mixture of non-bino
components, particularly Higgsino
components~\cite{BirkedalHansen:2002am,Chattopadhyay:2003yk,Baer:2006dz,Baer:2007xd,Bae:2007pa,Chattopadhyay:2009fr}.
This is because the SI cross-section is dominated by t-channel Higgs
exchange diagrams, the size of which are determined by the Higgsino
component of the LSP. An observable signal at CDMS-II therefore also
favors a relative light CP-even Higgs mass.

In Table~\ref{tbl:softhigh} we present five benchmark points which
exhibit event rates comparable to those recently reported by the
CDMS-II collaboration, computed by integrating~(\ref{ds29}) over the
range~(\ref{CDMSrange}). We arrived at these points by considering a
simple unified model with universal scalar soft mass $m_0$ and
trilinear coupling $A_0$ at a high-energy scale, here taken to be
the grand unified scale $\Lambda_{\UV} = 2 \times 10^{16} \GeV$. We
then introduced non-universal gaugino soft masses $M_1$, $M_2$ and
$M_3$ via
\begin{equation} M_i = m_{1/2} (1 + \delta_i) \label{gaugmass}
\end{equation}
where $m_{1/2}$ represents the typical scale of the soft gaugino
masses and we will consider only $\delta_2,\,\delta_3 \neq 0$. This
approach has been shown to give supersymmetric models compatible
with all direct and indirect constraints, but with relatively large
spin-independent cross-sections for the LSP
neutralino~\cite{Corsetti:2000yq,BirkedalHansen:2001is}. This
approach is similar in spirit to the exploration conducted
in~\cite{Baer:2006dz,Baer:2007xd}.

%===================== benchmark soft terms ==============================
\begin{table}[tb] \caption{\label{tbl:softhigh} Soft term parameters for selected
points for comparison to reported CDMS signal. All soft mass
parameters are given in units of GeV and all points were selected to
have $\mu > 0$.}
\begin{ruledtabular}
\begin{tabular}{|c|c|c|c|c|c|} \hline
Point & A & B & C & D & E \\
\hline \hline
$m_0$ & 750 & 500 & 350 & 350 & 350 \\
$m_{1/2}$ & 750 & 500 & 575 & 575 & 575 \\
$A_0$ & 370 & 270 & 150 & 150 & 150 \\
$\tan\beta$ & 25 & 15 & 30 & 30 & 30 \\
\cline{1-6}
$\delta_2$ & 0.65 & 0.62 & -0.6 & 0.82 & -0.47 \\
$\delta_3$ & -0.35 & -0.3 & -0.3 & -0.35 & -0.3 \\
\cline{1-6}
\end{tabular}
\end{ruledtabular}
\end{table}
%===============================================================

The gluino mass plays an important role in these models, albeit an
indirect one. As is well known, the gluino soft mass $M_3$ at the
high scale influences the eventual value of the $\mu$ parameter (as
determined by solving the electroweak symmetry breaking conditions
at the low energy scale) via renormalization group effects on the
running of the Higgs scalar soft
masses~\cite{Kane:1998im,Kane:2002ap,Horton:2009ed}.
The consistent choice of $\delta_3 < 0$ is therefore preferred as
this tends to decrease the value of the $\mu$ parameter
and hence an increase in the Higgsino content of the eventual LSP.

%===================== LSP and CDMS results ==============================
\begin{table}[tb] \caption{\label{tbl:cdms} Properties of the LSP and resulting
CDMS-II recoil rate for examples in Table~\ref{tbl:softhigh}. Masses
and LSP properties were computed using~{\tt SuSpect 2.41}, while scattering cross-sections and the expected number
of events were computed using default values in {\tt DarkSUSY
5.0.4}. These numbers carry uncertainties associated with certain
nuclear matrix elements, as mentioned in the text. Event rates on
germanium nuclei assume an effective exposure of 183.6 kg-days.}
\begin{ruledtabular}
\begin{tabular}{|c|c|c|c|c|c|} \hline
Point & A & B & C & D & E \\
\hline \hline
$m_{\chi_1^0}$ (GeV) & 138 & 190 & 175 & 112 & 230 \\
\cline{1-6}
B\% & 3.0\% & 70.2\% & 0.3\% & 5.4\% & 40.9\% \\
W\% & 0.4\% & 0.4\% & 95.8\% & 0.5\% & 53.0\% \\
H\% & 96.6\% & 29.4\% & 3.9\% & 94.1\% & 6.1\% \\
\cline{1-6}
$\sigma^{\rm SI}_{\chi p}\times 10^{45}$ (cm$^2$) & 11.9 & 44.4 & 41.3 & 35.3 & 74.8 \\
$N_{\rm Ge}$ (184 kg-days)  & 0.51 & 1.36 & 1.30 & 1.65 & 1.90 \\
\cline{1-6}
\end{tabular}
\end{ruledtabular}
\end{table}
%===============================================================

The relevant information for connecting these models to the reported
signal at CDMS-II is collected in Table~\ref{tbl:cdms}, where we
give the mass of the LSP neutralino, the spin-independent
cross-section for elastic scattering on a single proton, and the
number of events expected at CDMS-II over the energy range
in~(\ref{CDMSrange}) with an effective exposure of $0.3\times 612 =
183.6$ kg-days. We have deliberately chosen examples with a variety
of neutralino mass matrices, as evidenced by the eigenvector of the
lowest mass state. This LSP wavefunction composition is indicated by
the entries B\%, W\% and H\% which give the percentage of bino
eigenstate, wino eignesate and Higgsino eigenstate, respectively.
These quantities were computed using~{\tt SuSpect
2.41}~\cite{Djouadi:2002ze} and {\tt
DarkSUSY}.

Given the low statistics in the reported CDMS data we believe that
any one of these models could conceivably be giving rise to the
observed recoils. As mentioned above, these interaction rates are
sensitive to still uncertain input numbers associated with certain
nuclear matrix elements. For example, the package {\tt
MicrOMEGAs}~\cite{Belanger:2008sj} uses a different default value
for the $\pi\,N$ $\Sigma$-term and consequently predicts an event
rate on germanium roughly twice as large as the values give in
Table~\ref{tbl:cdms}. We have therefore chosen examples with $0.5
\leq N_{\rm Ge} \leq 2.0$ to be conservative in our estimates. It
should also be noted that recent calculations of the matrix element
$\lang n| s \bar{s} | n\rang$~\cite{Toussaint:2009pz} may imply a
much lower effective value for $\sigma^{\rm SI}_{\chi p}$, implying
a more challenging environment for interpreting the CDMS-II events
as neutralino scatters~\cite{Giedt:2009mr}.
Of course accumulating more statistics may allow for a crude
measurement of the recoil energy spectrum and thus shed more light
on the nature of the scattering particle.\\

%%%%%%%%%%%%%%%%%%%%%%%%%%%%%%%%%%%%%%%%%%%%%%%%%%%%%%%%%%%%%%%%%%
{\em Implications for Other Dark Matter Search Experiments.}

As we saw in the previous section, a variety of LSP configurations
can give rise to the signal seen at CDMS-II. Given the low
statistics in the reported data it is not altogether clear whether
the events seen truly indicate the presence of new physics.
Additional data collection is, of course, imperative to a firm
determination that we are indeed seeing evidence of dark matter
elastic scatters with an effective cross-section of the order
$\sigma \simeq 10^{-44}\,{\rm cm}^2$.

One crucial cross-check will be the next round of results reported
at xenon-based detectors. The spin-independent interaction rates of
neutralinos on germanium and on xenon are highly correlated.
Interestingly, the implied rate for neutralinos scattering on xenon
inferred from the CDMS-II results implies something on the order of
one to five events in an exposure of 316.4~kg-days at Xenon10. Ten
events were in fact observed in the region where the signal was
expected, but this was consistent with their background
estimation~\cite{Angle:2007uj}. A confirming signal from a
liquid-xenon based detector in the near future is therefore
imperative for the supersymmetric interpretation of the CDMS events.

%====== Other dark matter related observables ==============================
\begin{table}[tb] \caption{\label{tbl:dark} Predictions of the benchmark models
in Table~\ref{tbl:softhigh} for other dark matter search experiments
and the thermal relic density of neutralinos. All calculations were
performed using {\tt DarkSUSY 5.0.4}.}
\begin{ruledtabular}
\begin{tabular}{|c|c|c|c|c|c|} \hline
Point & A & B & C & D & E \\
\hline \hline
$R_{\rm Xe}$ (kg$^{-1}$-yr$^{-1}$) & 1.32 & 3.64 & 3.66 & 4.63 & 5.15 \\
$N_{\rm Xe}$ (Xenon10) & 1.14 & 3.16 & 3.17 & 4.01 & 4.47 \\
$N_{\rm Xe}$ (Xenon100) & 17.4 & 47.9 & 48.1 & 60.9 & 67.7 \\
\cline{1-6}
$(\Phi_{\rm int}/\bar{J})\times10^{15}$ (cm$^{-2}$ s$^{-1}$) & 196.8 & 9.27 & 1020.4 & 243.8 & 164.1 \\
$\Phi_{\gamma\gamma/}\Phi_{\gamma\,Z}$ & 0.52 & 0.20 & 0.33 & 0.57 & 0.31 \\
\cline{1-6}
$\Phi_{\mu}$ (km$^{-2}$ yr$^{-1}$) & 114.0 & 171.3 & 63.4 & 130.7 & 123.5 \\
\cline{1-6}
$\lang \sigma v\rang_{WW} \times 10^{25}$ (cm$^2$) & 1.9 & 0.1 & 23.8 & 2.0 & 4.5 \\
\cline{1-6}
$\Omega\,{\rm h}^2|_{\rm therm}$ & 0.005 & 0.112 & 0.001 & 0.006 & 0.003 \\
\cline{1-6}
\end{tabular}
\end{ruledtabular}
\end{table}
%===============================================================

To compute the rate of elastic scattering on xenon-based targets we
modify the procedure of the previous section and use an integration
range for~(\ref{ds29}) that better reflects the sensitivities of the
Xenon10 apparatus:
\begin{equation} 5\,{\rm keV} \leq E_{\rm recoil} \leq
25\,{\rm keV}\, . \label{xenonrange} \end{equation}
The event rates for our benchmark points in units of recoils/kg/year
are given in the first entry of Table~\ref{tbl:dark}. In the second
entry we give the expected number of signal events in 316.4~kg-days
at Xenon10. These points are clearly in a region where the
liquid-xenon experiments should be sensitive to a signal. In the
third entry we give the expected event rate at the Xenon100 upgrade
assuming an exposure of 80~kg $\times$ 60~days = 13.15~kg-years.
Assuming adequate rejection of background events all of these
candidate models should give clear dark matter signals in the
earliest stages of running at Xenon100.

Observation of a clear signal above background in liquid xenon-based
detectors would (a) provide some assurance that the effect seen at
CDMS-II is truly the result of dark matter scattering, and (b) give
some further support to the conjecture that this involves elastic
scattering of supersymmetric neutralinos. Even within the
supersymmetric framework, however, it is unlikely that such an
observation alone can distinguish between the various models
represented by the benchmarks in Table~\ref{tbl:softhigh}. Several
indirect detection experiments, however, may be sensitive to the
differences in the LSP wavefunction that distinguish these models.

The Fermi-GLAST satellite is currently surveying the flux of gamma
ray photons from the location of the galactic center. If some
component of this flux comes from the annihilation of LSP
neutralinos it may be possible to observe an increase in the flux
over estimated astrophysical backgrounds.
The calculation of the flux depends on the microscopic physics of
the neutralino and its interactions, but also very strongly on the
macroscopic physics of the halo profile assumed for the galaxy. The
latter is conveniently summarized by a single parameter $\bar
J\lp\Delta\Omega\rp$ where $\Delta\Omega$ represents the solid angle
resolution of the observatory. We consider the commonly adopted
profile of Navarro, Frank and White (NFW)~\cite{Navarro:1995iw}
which gives a value of $\bar J\lp \Delta\Omega \rp = 1.2644\times
10^4$ and assume an angular resolution of $\Delta\Omega=10^{-5}$~sr.
The value of the estimated gamma ray flux for photons in the energy
range $1\GeV \leq E_{\gamma} \leq 200\GeV$ is given in
Table~\ref{tbl:dark}. The expected flux is generally quite large,
suggesting that a signal should be expected above background at
Fermi/GLAST in the near future. But we emphasize that these
estimates can vary over many orders of magnitude if different
assumptions about the halo profile are employed.

In addition to the diffuse gamma rays coming from neutralino
annihilations one also expects monochromatic signals from
loop-induced diagrams in which two neutralinos annihilate into a
pair of photons ($\gamma\gamma$ line) or into a photon and a Z-boson
($\gamma\,Z$ line)~\cite{Bergstrom:1997fh,Bern:1997ng,Ullio:1997ke}.
It has been noted that the ratio of the observed gamma ray fluxes
from these two processes is correlated with the wave-function
composition of the LSP~\cite{Ullio:2001qk}. Furthermore, taking the
ratio of the fluxes eliminates much of the halo-model dependence
from the analysis of the signal. The ratio of the fluxes
$\Phi_{\gamma\gamma}/\Phi_{\gamma\,Z}$ is given in
Table~\ref{tbl:dark} for our various benchmark points. They differ
by as much as a factor of two. Observing these ratios, however, at
atmospheric Chernekov telescopes will require a favorable halo model
profile and excellent energy resolution on the measurement of these
high-energy photons.

More promising signals may come from the search for neutrinos
arising from the annihilation of LSP neutralinos in a location
closer to the earth, such as the center of the sun, where issues of
the galactic halo do not come into play. Again we expect relatively
large differences in the flux of neutrinos between wino-like and
Higgsino-like LSPs. Some fraction of the neutrinos can eventually
exit the sun and be detected in experiments such as
IceCube~\cite{Halzen:2006mq} via conversion of muon neutrinos into
muons. To calculate this rate we integrate the differential flux of
conversion muons from solar-born as well as earth-born neutrinos
over the energy range $50\GeV \leq E_{\mu} \leq 300 \GeV$, assuming
an angular resolution of 3~degrees. The nominal target area for
IceCube is 1~km$^2$, but the effective area for detection of
neutrinos via muon conversion is smaller and can be cast as a
function of the muon
energy~\cite{GonzalezGarcia:2005xw,Halzen:2005ar,Barger:2007xf}.
Previous observations based on a limited set of 22~strings of
photomultiplier tubes put a limit on the size of the muon flux at
approximately 300-400 muons/km$^2$/year. The fluxes for these
benchmark models all fall well below these limits, as shown in
Table~\ref{tbl:dark}, though well within reach of the next data run
at IceCube. Even with an exposure of only 0.2~km$^2$-years all of
these models will produce at least 10~signal events at IceCube.

Finally, we mention the implications for the observed excess in
positrons in the energy range 10-100 GeV recently measured by the
PAMELA experiment~\cite{Adriani:2008zr}. It is by now well
established that standard MSSM models can fit the signal from PAMELA
only with some difficulty. For models with a significant wino or
Higgsino content, the observed rate implies a thermally-averaged
annihilation cross-section in to $W^+ W^-$ final states on the order
of $\lang \sigma v\rang_{WW} \sim 10^{-24}$
cm$^3$/s~\cite{Cirelli:2008pk,Grajek:2008pg}. Such rates can be
achieved for some of the models in Table~\ref{tbl:dark} though not
for all.\footnote{See also the benchmark cases given
in~\cite{Feldman:2009wv}} Models which achieve this high
annihilation rate tend to also be cases in which neutralinos
annihilate very effectively in the early universe, resulting in a
diminished thermal relic density. We chose one model (Point~B) which
gives a thermal relic density, as computed using {\tt DarkSUSY},
which fits the measurement as inferred from the WMAP
three-year~\cite{Spergel:2006hy} and five-year
data~\cite{Komatsu:2008hk} which we take to be
\beq 0.0855 \leq \Ochi \leq 0.1189\, , \label{omegah2} \eeq
at the $2\sigma$ level. For the other cases it will clearly be
necessary to invoke non-thermal
production~\cite{Moroi:1999zb,Nagai:2007ud,Nagai:2008se,Gelmini:2006pw}
to achieve the appropriate relic density.\\

%%%%%%%%%%%%%%%%%%%%%%%%%%%%%%%%%%%%%%%%%%%%%%%%%%%%%%%%%%%%%%%%%%
{\em Implications for Observation of SUSY at the LHC.}

%====== Table of benchmark phys spectrum ==============================
\begin{table}[tb] \caption{\label{tbl:masses} Relevant SUSY mass
spectra and total production cross sections at the LHC, for the
benchmark models in Table~\ref{tbl:softhigh}. All masses are in
GeV.}
\begin{ruledtabular}
\begin{tabular}{|c|c|c|c|c|c|} \hline
Point & A & B & C & D & E \\
\hline \hline
$m_{\chi_1^0}$ & 138 & 190 & 175 & 112 & 230 \\
$m_{\chi_2^0}$ & 152 & 254 & 235 & 130 & 239 \\
$m_{\chi_3^0}$ & 326 & 261 & 505 & 252 & 504 \\
$m_{\chi_4^0}$ & 1008 & 663 & 513 & 846 & 515 \\
$m_{\chi_1^\pm}$ & 146 & 243 & 175 & 123 & 234 \\
$m_{\chi_2^\pm}$ & 1008 & 663 & 514 & 846 & 515 \\
\cline{1-6}
$m_{\tilde{g}}$ & 1156 & 847 & 952 & 890 & 951 \\
$m_{\tilde{t}_1}$ & 826 & 607 & 719 & 544 & 709 \\
$m_{\tilde{t}_2}$ & 1284 & 925 & 862 & 964 & 865 \\
$m_{\tilde{b}_1}$ & 1155 & 853 & 809 & 766 & 812 \\
$m_{\tilde{b}_2}$ & 1271 & 903 & 874 & 943 & 871 \\
\cline{1-6}
$m_{\tilde{\tau}_1}$ & 740 & 520 & 344 & 338 & 352 \\
$m_{\tilde{\tau}_2}$ & 1079 & 724 & 414 & 752 & 424 \\
$m_{h}$ & 115 & 112 & 113 & 114 & 113 \\
\cline{1-6}
$\sigma_{\SUSY}^{7\,{\rm TeV}}$ (pb) & 1.3 & 0.3 & 1.2 & 2.7 & 0.4 \\
$\sigma_{\SUSY}^{10\,{\rm TeV}}$ (pb) & 2.3 & 1.2 & 2.5 & 5.1 & 1.3 \\
$\sigma_{\SUSY}^{14\,{\rm TeV}}$ (pb) & 4.0 & 4.1 & 5.7 & 10.0 & 3.7 \\
\cline{1-6}
\end{tabular}
\end{ruledtabular}
\end{table}
%===============================================================

A common feature of the models listed in Table~\ref{tbl:softhigh} is
a general compression of the gaugino mass spectrum relative to
unified models such as mSUGRA. This is particularly true of the mass
gap $\Delta^{\pm} \equiv m_{\chi_1^{\pm}} - m_{\chi_1^0}$ which is
crucial to many of the standard supersymmetric search strategies at
the LHC, as well as having a large impact on the triggering
efficiencies for supersymmetric events. We give certain key physical
masses for the models of Table~\ref{tbl:softhigh} in
Table~\ref{tbl:masses} in units of GeV. All masses are computed from
the high-scale boundary conditions after renormalization group
evolution using {\tt SuSpect 2.41}.

To analyze the signatures of these models at the LHC we generated
events using PYTHIA followed by a detector simulation using
PGS4~\cite{pythiapgs}. Sparticle decays are calculated using {\tt SUSY-HIT}~\cite{Djouadi:2006bz}. Data sets representing 1~fb$^{-1}$ were
generated for center of mass energies $\sqrt{s} = 10\TeV$ and
$\sqrt{s}=14\TeV$ for each model. In addition we considered a
suitably-weighted sample of 1~fb$^{-1}$ Standard Model background
events, consisting of Drell-Yan, QCD dijet, $t\,\bar{t}$,
$b\,\bar{b}$, $W$/$Z$+jets and diboson production at each $\sqrt{s}$
value. Events were analyzed using level one (L1) triggers in PGS4,
designed to mimic the CMS trigger tables~\cite{Ball:2007zza}.
Object-level post-trigger cuts were also imposed. We require all
photons, electrons, muons and taus to have transverse momentum
$p_T\geq 10$ GeV and $|\eta|<2.4$ and we require hadronic jets to
satisfy $|\eta|<3$.

%====== LHC 10TeV Discovery Modes ==============================
\begin{table}[tb] \caption{\label{tbl:LHC} Number of signal events for standard discovery
channels at the LHC for $\sqrt{s}=10\TeV$ for the benchmark models
of Table~\ref{tbl:softhigh}. }
\begin{ruledtabular}
\begin{tabular}{|c|c|c|c|c|c|}
%\hline
\multicolumn{6}{c}{Numbers of Events} \\
\hline
Point & A & B & C & D & E \\
\hline \hline
Multijets & 16 & 68 & 72 & 114 & 57 \\
$1\ell$ + jets & 17 & 78 & 61 & 70 & 19 \\
OS $2\ell$ + jets & 0 & 14 & 2 & 12 & 1 \\
SS $2\ell$ + jets & 0 & 1 & 2 & 4 & 0 \\
$3\ell$ + jets & 0 & 4 & 0 & 4 & 0 \\
\cline{1-6}
\multicolumn{6}{c}{Significance $S/\sqrt{B}$} \\
\hline
Point & A & B & C & D & E \\
\hline \hline
Multijets & 2.0 & 8.4 & 8.9 & 14.0 & 7.0 \\
$1\ell$ + jets & 1.1 & 5.2 & 4.1 & 4.7 & 1.7 \\
OS $2\ell$ + jets & NA & 4.4 & 0.6 & 3.8 & 0.3 \\
SS $2\ell$ + jets & NA & 1.0 & 2.0 & 4.0 & NA \\
$3\ell$ + jets & NA & 2.9 & NA & 2.9 & NA \\
\cline{1-6}
\end{tabular}
\end{ruledtabular}
\end{table}
%===============================================================

The last entries of Table~\ref{tbl:masses} gives the total SUSY
production cross-section at the LHC for center of mass energies
of~7, 10~and~14~TeV for our benchmark points. Given the relatively
low cross-sections and reduction in leptonic signatures (as we
describe below) it is highly unlikely that these models will give
significant event rates above the Standard Model background in
500~pb$^{-1}$ or less integrated luminosity for $\sqrt{s} = 7\TeV$.
However, the prospects brighten considerably at higher energies,
particularly if one assumes that measurements of jet $p_T$ and
$\not\!\!{E_T}$ are reasonably reliable.

We begin with standard SUSY discovery modes~\cite{Baer:1995nq},
slightly modified to maximize the signal significance for these
models. These five signatures are collected in Table~\ref{tbl:LHC}
for 1~fb$^{-1}$ of integrated luminosity at $\sqrt{s} = 10\TeV$. In
all cases we require transverse sphericity $S_T \geq 0.2$ and at
least 250~GeV of $\not\!\!{E_T}$ except for the trilepton signature,
where only $\not\!\!{E_T} \geq 200 \,{\rm GeV}$ is required. The
multijet channel includes a veto on isolated leptons and requires at
least four jets with the transverse momenta of the four leading jets
satisfying $p_T \geq \left(200,150,50,50\right)$ GeV, respectively.
For the leptonic signatures we include only $e^\pm$ and $\mu^{\pm}$
final states and demand at least two jets with the leading jets
satisfying $p_T \geq \left(100,50\right)$ GeV, respectively.  We
also display results in Table~\ref{tbl:LHC2} for 1~fb$^{-1}$ of
integrated luminosity at $\sqrt{s} = 14\TeV$.

The multijet channel is promising already at 1~fb$^{-1}$ at
$\sqrt{s}=10\TeV$ for all models with the exception of point~A,
which has the largest gluino mass. Points which give a mass
difference $\Delta^{\pm} \gappeq 10\GeV$ are capable of giving hard
enough leptons to produce a reasonable signal significance in the
one-, and even two-lepton channels. The compressed spectrum tends to
result in fewer leptonic events due to the inability to trigger
effectively on leptons when their transverse momentum drops much
below 10~GeV. The reduction in leptonic signal is likely to be a
generic property of supersymmetric models that give rise to large
event rates at CDMS-II. On the other hand, all models should give
rise to detectable signals at $\sqrt{s} = 14\TeV$ in multiple
channels. We also note that all should have at least three states
which can be produced and observed at a $\sqrt{s}=500\GeV$
$e^+\,e^-$ linear collider. This is also likely to be a generic
property of models which fit the CDMS-II signal well.\\

%====== LHC 14TeV Discovery Modes ==============================
\begin{table}[tb] \caption{\label{tbl:LHC2} Number of signal events for standard discovery
channels at the LHC for $\sqrt{s}=14\TeV$ for the benchmark models
of Table~\ref{tbl:softhigh}. }
\begin{ruledtabular}
\begin{tabular}{|c|c|c|c|c|c|}
%\hline
\multicolumn{6}{c}{Numbers of Events} \\
\hline
Point & A & B & C & D & E \\
\hline \hline
Multijets & 99 & 321 & 402 & 436 & 298 \\
$1\ell$ + jets & 62 & 336 & 202 & 310 & 111 \\
OS $2\ell$ + jets & 8 & 45 & 12 & 45 & 7 \\
SS $2\ell$ + jets & 2 & 19 & 6 & 16 & 3 \\
$3\ell$ + jets & 3 & 8 & 4 & 6 & 1 \\
\cline{1-6}
\multicolumn{6}{c}{Significance $S/\sqrt{B}$} \\
\hline
Point & A & B & C & D & E \\
\hline \hline
Multijets & 6.6 & 21.4 & 26.9 & 29.1 & 19.9 \\
$1\ell$ + jets & 2.5 & 13.6 & 8.2 & 12.5 & 4.5 \\
OS $2\ell$ + jets & 1.3 & 7.4 & 2.0 & 7.4 & 1.2 \\
SS $2\ell$ + jets & 0.8 & 7.2 & 2.3 & 6.0 & 1.1 \\
$3\ell$ + jets & 1.2 & 3.3 & 1.6 & 2.5 & 0.4 \\
\cline{1-6}
\end{tabular}
\end{ruledtabular}
\end{table}
%===============================================================

%%%%%%%%%%%%%%%%%%%%%%%%%%%%%%%%%%%%%%%%%%%%%%%%%%%%%%%%%%%%%%%%%%
{\em Conclusions.}

If the events observed in the recent CDMS-II five-tower results are
interpreted as scattering of supersymmetric neutralinos, we may
begin to construct a picture of what to expect at future
experiments. It is highly likely, though by no means necessary, that
the soft supersymmetry breaking gaugino masses of the MSSM are
non-universal when computed at some high-energy input scale. The
CDMS-II results would favor a suppression of the gluino mass
relative to the prediction of models such as mSUGRA. They may also
imply a relatively compressed spectrum in the gaugino sector and
possible diminishment of lepton-based discovery channels at the LHC.
If a predominantly wino-like or Higgsino-like neutralino is the LSP
we may be forced to consider scenarios which give rise to
non-thermal production of relic neutralinos, such as those motivated
from string theory~\cite{Watson:2009hw}. In all scenarios which can
give rise to $\order(1)$ events in the reported effective exposure
at CDMS-II we should expect clear corroborating signals at the next
round of liquid-xenon based direct detection experiments as well as
clear signals at indirect detection experiments, most notably the
neutrino-induced muons at IceCube.

%%%%%%%%%%%%%%%%%%%%%%%%%%%%%%%%%%%%%%%%%%%%%%%%%%%%%%%%%%%%%%%%%%
{\em Acknowedgements.}

This work was supported by National Science Foundation Grant
PHY-0653587.

%\bibliography{omega}

\end{document}